\documentstyle[aps,multicol,epsf]{revtex}
\begin{document}
\draft
\date{\today}
\title{Method for Generating Long-Range Correlations for Large Systems}
\author{Hern\'an A. Makse$^{(a)}$,  Shlomo Havlin$^{(a),(b)}$, Moshe
Schwartz$^{(b)}$ and H. Eugene
Stanley$^{(a)}$}
\address{$^{(a)}$ Center for Polymer Studies and  Department of Physics,
Boston University,  Boston, Massachussetts 02215  }
\address{$^{(b)}$Department of Physics, Bar-Ilan University,
Ramat-Gan, ISRAEL}

\maketitle
\begin{abstract}
We propose a new method
to generate a sequence of random numbers with long-range
power-law correlations that overcomes known difficulties associated with
{\it large systems}.
The new method presents an improvement
on the commonly-used methods.
We apply the algorithm to
generate enhanced diffusion,
isotropic and anisotropic self-affine surfaces,
and isotropic and anisotropic correlated percolation.
\end{abstract}
\pacs{PACS numbers:}

\begin{multicols}{2}
\narrowtext

\section{Introduction}

Recently, the study of physical  systems displaying long-range power-law
correlation has attracted considerable attention.
Long-range correlations have been found in a wide number of systems
including  biological, physical, economical,
geological and urban systems \cite{review}.
Attempts to study and characterize such systems are often based on
numerical methods to generate correlated noise \cite{ffm,other}.
One of the most used methods to generate a sequence of random
numbers with power-law correlations is the Fourier filtering method
(Ffm) \cite{ffm,ck,sona}.
It consists of  filtering the Fourier components of a
uncorrelated sequence of random numbers with a suitable power-law filter
in order to  introduce correlations among the variables.
This method has the disadvantage of presenting a finite cutoff in the
range over which the variables are actually correlated \cite{ck,sona,feder2}.
As a consequence, one must generate a very large sequence of
numbers, and then use only the small fraction of them that are actually
correlated (this fraction can be as
small as $0.1\%$ of the initial length of the sequence \cite{ck,sona}).
This limitation makes the Ffm not suitable for the
study of scaling properties in the limit of large systems.

Here we modify the  Ffm in
order to remove the cutoff in the range of correlations.
We show
that in the modified method the actual correlations extend to the
{\it whole system}. We also apply
the method to generate several systems such as fractional Brownian motion,
self-affine surfaces and  long-range correlated percolation.

\section{Fourier filtering method }

We  start by defining the  Ffm \cite{ffm,ck,sona} for the $d=1$ case ($d$
is the dimension of the sample).
Consider a  stationary sequence of $L$ uncorrelated
random numbers
 $ \{ u_{i} \}_{i=1,..,L}  $.
The  correlation function is
$ \langle  u_{i} ~  u_{i+\ell} \rangle \sim \delta_{\ell,0}$,
with $\delta_{\ell,0}$ the Kronecker delta, and
the brackets denote an
average with respect to a Gaussian distribution.
The goal is to use the sequence
 $ \{ u_{i} \}  $, in order to  generate a new
sequence, $ \{  \eta_{i} \}$, with a long-range power-law
correlation function $C(\ell)$  of the form
\begin{equation}
\label{coreal}
 C(\ell) \equiv  \langle  \eta_{i} ~  \eta_{i+\ell} \rangle \sim
\ell^{-\gamma}  \qquad [\ell \to \infty].
\end{equation}
Here, $\gamma$ is the correlation exponent, and the long-range
correlations are relevant for  $0<\gamma<d$,
where $d=1$.
The spectral density $S(q)$ defined as the Fourier transform of
$C(\ell)$ \cite{reif} has the asymptotic form

\begin{equation}
\label{cofourier}
S(q)  =  \langle  \eta_{q} ~  \eta_{-q} \rangle
\sim q^{\gamma-1}  \qquad [q \to 0].
\end{equation}
Here $ \{ \eta_{q} \}$ correspond to
the Fourier transform coefficients of
$\{ \eta_{j} \}$, and  satisfies
\begin{equation}
\label{fourierg}
\eta_q = \left(S(q)\right)^{1/2} ~ u_q ,
\end{equation}
where $\{ u_q\}$ are the Fourier transform coefficients of $\{ u_{i} \}$.

The actual numerical algorithm for Ffm consists of the following steps:

{\bf \it (i)} Generate a one-dimensional sequence $ \{ u_{i} \} $
of uncorrelated random numbers with a Gaussian distribution,
and calculate the Fourier transform coefficients $\{u_q\}$ \cite{practice}.

{\bf \it (ii)} Obtain $\{\eta_q\}$ using (\ref{cofourier})
and (\ref{fourierg}).

{\bf \it (iii)} Calculate the inverse Fourier transform of
$\{ \eta_q \}$ to obtain $\{\eta_i\}$, the sequence
in real space with the desired
power-law correlation function (\ref{coreal}).

\section{Present method}
The Ffm method has
been applied in a number of studies of correlated systems
\cite{review,ck,sona}. However, an analysis
of the method for large $L$ shows that,
by following the above procedure, one ends up
with a sequence of correlated numbers whose range of correlations, for
$d=1$, is only about $0.1\%$  of the system size.
For example, from an initial
sequence of $10^6$ numbers, only $10^3$
numbers show the desired power-law correlations \cite{ck}.
For $d=2$, the range of correlations increases to
$1\%$  of the system size \cite{sona}.
In order to remove this artificial cutoff in the correlations,
we modify the Ffm algorithm as follows:

{\bf \it (a)}  To calculate the spectral density $S(q)$,
a well-defined correlation function in the real space
is needed.
The function $ C(\ell) = \ell^{-\gamma}$ has a singularity
at $\ell=0$. We replace (\ref{coreal}) with a
slightly modified correlation function
that has the desired power-law behavior for large $\ell$,
 and is well-defined at the origin
\begin{equation}
\label{coreal2}
 C(\ell)  \equiv (1+\ell^2)^{-\gamma/2}.
\end{equation}

{\bf \it (b)} The relation (\ref{fourierg}) is based on the
convolution theorem, and therefore   the desired correlation function
(\ref{coreal2}) must satisfy the proper  periodic boundary condition.
The function $C(\ell)$ can be naturally extended to
negative values of $\ell$ due to the $\ell^2$ dependence.
We define
(\ref{coreal2})  in the interval
$[-L/2,..., L/2]$,  and impose
periodic boundary conditions, i.e. $C(\ell)=C(\ell+L)$.

{\bf \it (c)} The discrete Fourier transform of
(\ref{coreal2})--- needed to obtain
$\eta_q$ using (\ref{fourierg})--- can now be calculated
analytically,
\begin{equation}
\label{analfourier2}
S(q) = \frac{2 \pi^{1/2}}{ \Gamma(\beta+1)} ~
\left(\frac{q}{2}\right)^{\beta}
 ~ K_\beta(q),
\end{equation}
where $q$ takes values $q=2\pi m/L$
with $m = -L/2,..., L/2$,
 $K_\beta(q)$
is the modified Bessel function of order $\beta=(\gamma-1)/2$, and
$\Gamma$ is the Gamma function.

The modified Bessel functions satisfy the asymptotic relations
\begin{equation}
\label{bessel}
K_{\beta}(q) =   \left\{ \begin{array}{ll}
                           \frac{\Gamma(\beta)}{2} ~ (\frac{q}{2})^\beta &
							 \mbox{if $q \ll 1$} \\
                          \sqrt{\frac{\pi}{2 q}} ~ e^{-q}     &
							\mbox{if $q \gg 1$,}
                                 \end{array}
                        \right.
\end{equation}
for $\beta$ positive and by definition $K_{-\beta}=K_{\beta}$.
Then for small values of $q$,
 (\ref{analfourier2})
gives the same asymptotic form as (\ref{cofourier}).
However, the Bessel function introduces a cutoff for  large $q$ in the
sense that  $S(q)$ has a faster exponential decay. This cutoff,
while irrelevant to the long-distance scaling, is
very important for the validity of the whole Fourier analysis because
it avoids aliasing effects (see Chapter $7$ in \cite{recipes}). The
cutoff in the Fourier space is thus responsible for eliminating the
cutoff in real space observed in the Ffm method.

In order to perform the above steps
numerically,
we employ the Fast Fourier transform \cite{recipes,fast}.
Due to the periodic
boundary condition imposed on the correlation function,
it follows that the correlated sample satisfies the same periodicity. If
one requires a sequence with open boundary conditions, we generate
twice as many
numbers  and then split the sequence in two parts
\cite{numerical}.

To test the actual
correlations  of the generated sample $\{ \eta_i \}$
we calculate $C(\ell)$
averaging
over different realizations of
random numbers.
Figure \ref{1d}
shows a plot of the actual correlations obtained for different
values of $\gamma$ and for a sequence of $L=2^{21}$ numbers.
It is seen that the
long-range correlations exist for the {\it whole} system.
The nominal values of $\gamma$ obtained
from the best fits are also the same, within the error bars, as the desired
input values.

To summarize the method, the correlation function we propose
is well-defined and  satisfies
the correct power-law behavior in the real space. Its Fourier
transform has  the correct power law  at small
frequencies, and  presents a cutoff for large frequencies that avoids
aliasing effects, and leads to the  infinite
long-range behavior in  real space. An alternative method in which
$S(q)$ was calculated numericaly, in contrast to the analytical
expression (\ref{analfourier2}), is given in \cite{makse}.

\section{Applications}

In the following we apply the proposed method
to several physical problems.
\subsection{Generating fractional Brownian motion (fBm).}
We map the variables $\{ \eta_i\}$ onto the steps of
a correlated random walk, and
define the position of the walker at step $t$ by
$x(t) = \sum_{i=1}^{t} \eta_i$.
Then, $x(t)$
corresponds to a
$t$-step fBm, and the sequence of increments  $
\{\eta_i\}$ is called fractional Gaussian noise (fGn) \cite{mandelbrot}.
An important quantity is the mean-square displacement of the fBm
whose asymptotic behavior is
\begin{equation}
\label{msd}
\langle |x(t)-x(t_0)|^2  \rangle \sim
|t-t_0|^{2-\gamma}.
\end{equation}

Thus the long-range correlations
lead to enhanced diffusion \cite{shlesinger}
$  \langle |x(t)-x(t_0)|^2 \rangle \sim |t-t_0|^{2 H}  $ for $0 < \gamma < 1$
with $H = 1 - \gamma / 2$ \cite{antipersistent}.
Figure \ref{fbm}
shows the plots of the mean square displacement
for different degree of correlations. The fits confirm the
validity of the long range correlations among the variables in the whole
system size.

\subsection{ Generating long-range correlations in $d$ dimensions.}
The algorithm can be easily generalized to higher dimensions.
In a $d$-dimensional cube of volume $L^d$
the desired correlation function takes the form
\begin{equation}
\label{coreal2d}
C(\vec{\ell}) = \left(1+\sum_{i=1}^d \ell_i^2\right)^{-\gamma/2},
\end{equation}
with the corresponding periodic boundary condition,
$C(\vec{\ell})=C(\vec{\ell}+\vec{L})$.
The spectral density is
\begin{equation}
\label{analfourier2d}
S(\vec{q}) = \frac{2\pi^{d/2}}{\Gamma(\beta_d+1)} ~
\left(\frac{q}{2}\right)^{\beta_d} ~  K_{\beta_d}(q),
\end{equation}
where $q=|\vec{q}|$,
$q_i=2\pi m_i/L$,
$-L/2 \leq m_i \leq L/2$, $i=1,...,d$, and
$\beta_d=(\gamma-d)/2$.
In the two-dimensional case the correlated variables are defined in a xy
square lattice
$\{\eta_{i , j}\}$. Figure \ref{2d} shows a
test of the actual correlations
obtained in two dimensions for different degree of correlations, and for a
system of  $L=2^{11}$.

\subsection{ Generating fBm in two dimensions}
The two-dimensional correlated numbers
$\{\eta_{i,j}\}$ can be used to generate
two-dimensional fBm. We propose the following definition
\cite{kent}
\begin{equation}
\label{prescription}
h(t,s) \equiv \sum_{i=1}^{t}
\eta_{i,s} + \sum_{j=1}^{s}
\eta_{t,j} .
\end{equation}

After some algebra, we find that
when the numbers $\{\eta_{i,j}\}$ are long-range
correlated then
\begin{equation}
\label{scalingwidth}
\langle |h(t,s)-h(t_0,s_0)|^2 \rangle  \sim
|(t-t_0)^2+(s-s_0)^2|^{1-\gamma/2}
\end{equation}
Thus, using the correlated numbers
with $0 < \gamma < 2$, fBm can be generated with
exponent given by $H = 1-\gamma/2$ and $0 < H < 1$. A landscape with
this scaling behavior is also called self-affine surface \cite{vicsek}.

\subsection{Generating anisotropic long-range correlations}
Many physical systems display not only correlations but also anisotropy
\cite{review}
reflected in different correlation exponents along different
directions. We generalize the algorithm for this case.
We propose a correlation function suitable for
two-dimensional anisotropic systems
\begin{equation}
C(r,\varphi) = r^{-\gamma_x} \cos^2 \varphi +
r^{-\gamma_y} \sin^2 \varphi,
\label{anico}
\end{equation}
where $(r,\varphi)$ are the polar coordinates.
The spectral density is

\begin{equation}
\begin{array}{cccc}
\displaystyle{S(q, \varphi_q)} &\displaystyle{=}& \displaystyle{ \frac{
\pi^{3/2} \Gamma(1+\beta_x/2)}{2^{1-\beta_x}\Gamma(2-\beta_x/2)}
\frac{\cos^2\varphi_q}{q^{\beta_x}}}& \displaystyle{+}\\
& & \displaystyle{
\frac{
\pi^{3/2} \Gamma(1+\beta_y/2)}{2^{1-\beta_y}\Gamma(2-\beta_y/2)}
\frac{\sin^2\varphi_q}{q^{\beta_y}}}&,
\end{array}
\label{anifo}
\end{equation}
with $\beta_x = 2-\gamma_x$, and $\beta_y=2-\gamma_y$.
Then the proposed method can be applied to generate anisotropic correlated
numbers.
This method might be suitable for the
simulation of geological reservoir systems for which strong
anisotropy  is found \cite{review}.
Moreover, after generating the anisotropic variables $\eta$, we
can apply the procedure outlined in {\bf C} in order to obtain an
anisotropic self-affine surface.

\subsection{The correlated percolation problem}
A qualitative check of the impact of long-range correlations
for physical systems can be obtained
by applying the proposed method to  a concrete physical problem: the
correlated percolation \cite{weinrib}.
The properties of long-range correlated site percolation in the
square lattice have been recently studied \cite{sona}.
However,
these studies were limited to systems not larger than $104 \times 104$ sites.
The method we present here allows us to study this
problem in the limit of large systems.
Figure \ref{percolation} illustrates the results obtained for
site percolation on a square lattice of $1024 \times
1024$ sites.
We see that the introduction of long-range correlations among the
occupancy variables strongly affects the morphology of the system. In the
correlated case the clusters look more compact than in the uncorrelated
case. The lack of correlations in the uncorrelated case is seen from the
presence of many small black holes inside the large clusters (Fig.
\ref{percolation}d).
Also, at small concentrations there are only small clusters (Fig.
\ref{percolation}c)
while for the correlated case, large clusters are present even at   low
concentration (Fig.  \ref{percolation}a).

We wish to thank
R. Cuerno, P. Jensen, P. R. King
C.-K. Peng, S. Prakash, R. Sadr, and S. Tomassone for
discussions. We thank K. B. Lauritsen for discussions leading to
(\ref{prescription}). The authors thank BP and NSF for financial support.

\begin{figure}
%\centerline{
%\vbox{ \hbox{\epsfxsize=7cm \epsfbox{fig1.eps}}}
%}
\caption{ A log-log plot of the  average correlation $C(\ell)$ of $50$
correlated samples obtained with the proposed method
for $L = 2^{21}$.
Shown are results for
different values of the desired $\gamma$ = 0.2, 0.4,
0.6, and 0.8 (from top to bottom). The dashed lines represent the
best fits which yield the nominal values of $\gamma =$ $0.19\pm0.02$,
$0.39\pm0.02$, $0.60\pm0.03$ and $0.79\pm0.03$. The correlations are
calculated until $L/2$ due to the periodic boundary conditions.}
\label{1d}
\end{figure}

\begin{figure}
%\centerline{
%\vbox{ \hbox{\epsfxsize=7cm \epsfbox{fig2.eps}}}
%}
\caption{Log-log plot of the mean square displacement
for the fBm, for the same values of the
desired $\gamma$ as in Fig. $1$ (from bottom to top). The slopes
of the linear fits yield $2-\gamma=$ $1.78\pm0.02$, $1.60\pm0.02$,
$1.42\pm0.03$,  and  $1.23\pm0.03$
respectively,
in agreement with
$(7)$.}
\label{fbm}
\end{figure}

\begin{figure}
%\centerline{
%\vbox{ \hbox{\epsfxsize=7cm \epsfbox{fig3.eps}}}
%}
\caption{Log-log plot of the correlations along the diagonal
direction in a square lattice of
$2^{11} \times 2^{11}$.
Shown are results for different values of $\gamma = $0.4, 0.8, 1.2 and
1.6 (from top to bottom),
and we take averages over
$50$ samples. The fits yield
nominal values of $\gamma = 0.41\pm0.02$,
$0.81\pm0.03$, $1.20\pm0.03$ and $1.59\pm0.04$.}
\label{2d}
\end{figure}

\begin{figure}
%\centerline{
%$(a)$
%\vbox{ \hbox{\epsfxsize=3cm \epsfbox{perco-10-0.2-0.2-no-per.ps}
%       \hspace*{0.5cm} $(b)$  \epsfxsize=3cm
%        \epsfbox{perco-10-0.2-pc-no-per.ps}}}
%           }
%\vspace*{0.0cm}
%\centerline{
%$(c)$
%\vbox{ \hbox{\epsfxsize=3cm \epsfbox{perco-10-unco-0.2-no-per.ps}
%       \hspace*{0.5cm} $(d)$  \epsfxsize=3cm
%        \epsfbox{perco-10-unco-pc-no-per.ps}}}
%           }
%\centerline{
%$(e)$
%\vbox{ \hbox{\epsfxsize=3cm \epsfbox{bessel-10-0.2-1.8-0.2.ps}}
%        }
%}
\caption{Site percolation in the square lattice of $1024 \times 1024$
for different degrees of correlations and  concentrations.
Figures
$4a$ and $4b$ correspond to the correlated case with $\gamma = 0.2$, while
Figs. $4c$ and $4d$
correspond to the uncorrelated percolation problem, below and at the
threshold for both cases, respectively.
Unoccupied sites are in black and occupied sites are in white. Figure
$4e$ shows the case of anysotropic percolation generated with the method
of {\bf D} for $\gamma_x=0.2$ and
$\gamma_y=1.8$, and for the same
concentration as in Fig. $4a$.
We notice how the
clusters are elongated along the direction of the smaller exponent
(strong correlations).  The
figures are generated using the same seed for the random number generator.}
\label{percolation}
\end{figure}

\end{multicols}


\vspace*{-0.5cm}

\begin{references}

\vspace*{-0.5cm}

\bibitem{review}
{\em Dynamics of Fractal Surfaces}, edited by F. Family and
T. Vicsek (World Scientific,
Singapore, 1991);
M. Sahimi, Rev. Mod. Phys. {\bf 65}, 1393 (1993);
A. Bunde, S. Havlin, eds., {\it Fractals
in Science} (Springer-Verlag, Berlin 1994);
S. Havlin {\it et al.} Phys. Rev. Lett. {\bf 61}, 1438 (1988); H. A.
Makse, S. Havlin, and H. E. Stanley, Nature {\bf 377}, 608 (1995).


\bibitem{ffm}
D. Saupe in  {\it The Science of
Fractal Images}, H.-O. Peitgen and D. Saupe, eds.
(Springer-Verlag, New York 1988);
J. Feder, {\it Fractals} (Plenum Press, New York 1988).

\bibitem{other}
See e.g.  B. B. Mandelbrot, Water Resour. Res. {\bf 7},
543 (1971);
R. F. Voss in {\it Fundamental Algorithms in Computer
Graphics}, edited by R. A. Earnshaw  (Springer-Verlag, Berlin 1985).



\bibitem{ck}
C.-K. Peng {\it et al.} Phys. Rev. A {\bf 44}, 2239 (1991).

\bibitem{sona}
S. Prakash {\it et al.} {\it ibid} {\bf 46}, R1724 (1992).

\bibitem{feder2}
Other methods present similar problems. See for instance Chapter 9 in
Feder's book
\cite{ffm}.

\bibitem{reif}
R. Reif, {\it Fundamental Statistical and Thermal Physics} (Mc Graw-Hill,
1965).

\bibitem{practice}
In practice one can generate directly $\{u_q\}$ from a sequence of
uncorrelated random numbers.


\bibitem{recipes}
W. H. Press, S. A. Teukolsky, W. T. Vetterling, and B. P. Flannery,
{\it Numerical Recipes in Fortran}, 2nd ed. (Cambridge University Press,
Cambridge  1992).

\bibitem{fast}
The calculation of the regular Fourier transform involves  $O(L^2)$
operations. Using the Fast Fourier
transform algorithm \cite{recipes}
the process is computed in $O(L~\log L)$
operations, which gives a drastic difference in computing time, and
makes this method very fast.


\bibitem{numerical}
Another  numerical detail that should be considered
is that the correlation
function in the Fourier space in not defined for  $q=0$.
This comes from the fact that we are
using a continuum limit to calculate the Fourier transform instead of
the discrete definition. However,
the zero frequency  only adds an additive
constant to the numbers,
and does not affect the scaling properties
of the sequence. This singularity
can be avoided anyway, by assigning a suitable numerical value $0 < m_0 < 1$
instead of $m=0$.


\bibitem{makse}
H. A. Makse,  S. Havlin, H. E.
Stanley, and M. Schwartz,
Proc. 1993 Int. Conf. Complex Systems in
Computational Physics, Buenos Aires,
Chaos, Solitons and
Fractals {\bf 6}, 295 (1995). Independently, the same method was also
published in N.-N.
Pang, Y.-K. Yu,  and
T. Halpin-Healy, Phys. Rev. E, {\bf 52} 3224 (1995).


\bibitem{mandelbrot}
B. B. Mandelbrot and J. W. Van Ness, SIAM Rev. {\bf 10}, 442 (1968).

\bibitem{shlesinger}
M. F. Shlesinger, J. Klafter, Phys. Rev. Lett. {\bf 54}, 2551 (1985).

\bibitem{antipersistent}
Antipersistent behavior $(0<H<1/2)$ cannot be obtained using
(\ref{coreal2}). In this case, the correlation function must satisfy
$C(\ell)\sim-\ell^{-\gamma}$, for $\ell\to\infty$, and $\int_0^\infty
C(\ell) d\ell = 0$ \cite{mandelbrot}.

\bibitem{kent}
See also K. B. Lauritsen, Ph.D. thesis, Aarhus University (unpublished, 1994).

\bibitem{vicsek}
T. Vicsek, {\it Fractal Growth Phenomena},  2nd ed.
(World Scientific, Singapore 1991).

\bibitem{weinrib}
A. Coniglio, {\it et al.}
J. Phys. A {\bf 10}, 205-209 (1977);
A. Weinrib, {\it Phys. Rev. B} {\bf 29}, 387-395 (1984).




\end{references}
\end{document}